\documentclass[fleqn,usenatbib]{mnras}

\usepackage{newtxtext,newtxmath}

\usepackage[T1]{fontenc}
\usepackage{ae,aecompl}

\usepackage{graphicx}	
\usepackage{amsmath} 
\usepackage[utf8]{inputenc}

\usepackage{aas_macros}
\usepackage{bm} 
\usepackage{fullpage}
\usepackage{graphicx}
\usepackage{multirow}
\usepackage{natbib}
\usepackage{subcaption}
\usepackage{siunitx}
   
\newcommand{\lcdm}{\ensuremath{\Lambda}CDM}

\newcommand{\Om}{\ensuremath{Om}}
\newcommand{\Omm}{\ensuremath{\Omega_\text{m}}}
\newcommand{\Omde}{\ensuremath{\Omega_\text{DE}}}

\newcommand{\Oml}{\ensuremath{\Omega_\Lambda}}
\newcommand{\diff} {\mathrm{\ensuremath{d}}} 
 
\newcommand{\deriv}[2]{\ensuremath{\frac{\diff {#1}}{\diff {#2}}}}

\newcommand{\fseight}{\ensuremath{f\sigma_8}}

\newcommand{\tens}[1]{\ensuremath{\mathbfss{#1}}}

\newcommand{\dlum}{\ensuremath{d_\text{L}}}

\newcommand{\chsq}{\ensuremath{\chi^2}}
\newcommand{\hMpc}[1]{\ensuremath{{#1}\,h^{-1}\,\mathrm{Mpc}}}

\title{Model-independent cosmological constraints from growth and expansion} 

\author{Benjamin
  L'Huillier,$^{1}$\thanks{E-mail:   benjamin@kasi.re.kr, shafieloo@kasi.re.kr}   
  Arman Shafieloo,$^{1,2}$ and
  Hyungjin
  Kim$^3$
\\
$^{1}$Korea   Astronomy   and   Space   Science   Institute,
  Yuseong-gu, 776 Daedeok daero, Daejeon 34055, Korea\\ 
$^{2}$University of  Science and Technology,  Yuseong-gu 217
Gajeong-ro, Daejeon 34113, Korea\\
$^{3}$Department  of  Applied  Mathematics,  University  of  Waterloo,
  Waterloo, Ontario, N2L 3G1, Canada
}

\date{Accepted February 6, 2018. Received February 4, 2018; in original form January 3, 2018}

\pubyear{2018}  

\begin{document}
\label{firstpage}
\pagerange{\pageref{firstpage}--\pageref{lastpage}}
\maketitle

\begin{abstract}
  Reconstructing the expansion history of the Universe from type Ia supernovae data, we
  fit  the   growth  rate   measurements  and   put  model-independent
  constraints on some key cosmological parameters, namely, $\Omm,\gamma$, and $\sigma_8$. 
  The  constraints  are  consistent  with those from the concordance  model within the framework of
  general  relativity, but  the current  quality  of the  data is  not
  sufficient to rule out modified gravity models.
  Adding the  condition that  dark energy  density should be  positive at all redshifts, 
  independently of its equation of state,  further constrains the parameters and interestingly 
  supports the concordance model.
\end{abstract}




\section{Introduction} 

The discovery  of the  acceleration of the  expansion of  the Universe
\citep{1998AJ....116.1009R, 1999ApJ...517..565P}  led to  the emergence
of the  \lcdm\ paradigm, further supported by the
study      of      the     cosmological      microwave      background
\citep{2003ApJS..148....1B, 2016A&A...594A..13P}  and  the  large-scale structures  
of  the Universe \citep[e.g.,][]{2005ApJ...633..560E,2017MNRAS.470.2617A}.
In this paradigm, gravity is  described by general relativity (GR), and the
energy  budget is  dominated  by the cosmological constant as dark  energy (DE), 
responsible for the acceleration of the expansion, and a
smooth, cold dark matter component. 
However, the nature  of DE is  one of the biggest  mysteries of modern
physics, and the simplest  candidate, the  cosmological constant,  poses theoretical
problems \citep[e.g.,][]{1989RvMP...61....1W,2003RvMP...75..559P}.
Alternatively, general  relativity may  not be  the correct  theory to
describe gravity, and the acceleration may reflect departure from GR.

At the  background level, for  a flat  universe, the expansion  of the
smooth Universe $h(z)=H(z)/H_0$ follows
\begin{align}
  h^2(z) &= \Omm (1+z)^3 + \Omde(z),
\end{align}
where $H_0$  is the  Hubble constant today,  $\Omm$ the  matter energy
density today,
\begin{equation}
  \label{eq:h}
  \Omde(z)  =  (1-\Omm) \exp\left(3\int_0^z \frac {1+w(z')} {1+z'}
  \diff z'\right), 
\end{equation}
the DE contribution to the energy density, and
$w(z) =  P_\mathrm{DE}/\rho_\text{DE}$ is the DE equation of
state.  
For  a cosmological  constant $\Lambda$,  $w\equiv-1$ and  $\Omde(z) =
\Oml \equiv 1-\Omm$, but current data do not rule out models such as 
quintessence or dynamical DE models \citep[e.g.][]{2017arXiv171003271O,2017NatAs...1..627Z}.

Meanwhile, at the perturbation level, the growth rate $f$ is defined as 
\begin{align}
  f(z) &= \deriv {\ln \delta} {\ln a} \simeq \Omm^\gamma(z),
\end{align}
where  $\gamma$   is  the  growth   index  \citep{2005PhRvD..72d3529L,
  2007APh....28..481L, 2017APh....86...41L}, and
\begin{align}
  \label{eq:Omz}
  \Omm(z) & = \frac{\Omm(1+z)^3}{h^2(z)}  
\end{align}
is the matter contribution to the energy density at a given redshift.

Observationally, redshift-space distortion measures
\begin{align}
  \label{eq:fs8}
  \fseight(z) & \simeq \sigma_8(0) \Omm^\gamma(z)
  \exp{\left(-\int_0^z {\Omm^\gamma(z')
      \frac{\diff z'}{1+z'}}\right)},
\end{align}
where $\sigma^2_8(z)$ is the mass variance in a \hMpc{8} sphere. 
For simplicity, we will denote $\sigma_8=\sigma_8(0)$ when there is no
ambiguity.

From   eq.~\eqref{eq:Omz}  and~\eqref{eq:fs8},   it   is  clear   that
\fseight\ depends on $(\Omm,\gamma,\sigma_8)$ as well as the expansion
history $h(z)$.
In  general  relativity  (GR),   $\gamma\simeq  0.55$,  while  modified
theories of gravity such  as $f(R)$ \citep{2010LRR....13....3D} or DGP
\citep{2000PhLB..485..208D}      predict      different      (possibly
scale-dependent) values of $\gamma$ \citep{2007APh....28..481L}.  
Therefore, $\fseight$ is a powerful probe of gravity.
Moreover, joined measurements of $h(z)$ and \fseight\ can help
break degeneracies  between modified gravity theories  and dark energy
\citep{2005PhRvD..72d3529L,2017APh....86...41L}. 
Therefore, it has been used to test the \lcdm\ model or alternative gravitys theories \citep[e.g.,][]{2008PhRvD..77b3504N, 2010PhRvD..81h3534B, 2012IJMPD..2150064B, 2013PhRvD..87b3520S, 2015JCAP...01..004G, 2015PhRvD..91f3009R, 2016arXiv161200812M, 2017PhRvD..96b3542N, 2017MPLA...3250054S}.

In  this paper,  we aim  to constrain  some key  cosmological  parameters, namely, $\Omm, \sigma_8$, and $\gamma$,  by
fitting the  growth data  using model-independent  expansion histories
that do not assume any DE model. 

\S~\ref{sec:method} describes  the data  and method, our  results are
shown  in   \S~\ref{sec:res}. \S~\ref{sec:pos} explores the effects of
restricting the DE density to be positive at all redshift,
and   our  conclusions  are   drawn  in \S~\ref{sec:ccl}. 

\section{Method}
\label{sec:method}

We used  reconstructed expansion  histories from the  Joint Lightcurve
Analysis  \citep[JLA,][]{2014A&A...568A..22B} and  combined them  with
growth measurements.

\subsection{Model-independent reconstructions of the expansion history}

We  reconstructed  the  expansion  history from  the  JLA  compilation (unbinned data with full covariance matrix)
using     the     iterative    model-independent     smoothing     method
\citep{2006MNRAS.366.1081S, 2007MNRAS.380.1573S, 2017JCAP...01..015L}.  
Starting  from some  initial  guess $\hat\mu_0(z)$,  we calculate  the
smooth distance modulus at any redshift $z$ at iteration $n+1$ as
\begin{multline}
  \label{eq:smooth}
  \hat\mu_{n+1}(z) = \hat\mu_n(z) \\
  + N(z)
  \sum_i{    \frac{  \mu(z_i)   -  \hat  \mu_n(z_i)}  {\sigma_i^2}
    \exp{ \left( - \frac{ \ln^2 \left(  \frac{ 1+z_i } {1+z} \right) }
      {2 \Delta^2} \right)}, }
\end{multline}
where
\begin{align}
  N^{-1}(z)  &=  \sum_i{  \frac  1  {\sigma_i^2} \exp{  \left(  -
               \frac{  \ln^2  \left(  \frac{1+z_i} {1+z}  \right)}  {2
                 \Delta^2} \right)} }
\end{align}
is a normalization factor, $\mu(z_i)$  and $\sigma_i$ are the measured
distance  modulus and  its  associated error  at  redshift $z_i$,  and
$\Delta = 0.3$ is the smoothing length.

We then obtain the smooth luminosity distances
\begin{align}
  \dlum(z) & = 10^{\mu/5-5}\si{Mpc}.
  \intertext{Assuming a flat universe, we can calculate }
  h(z) & = \frac c {H_0} \left[\deriv {} {z} \left( \frac{ \dlum(z)}
  {1+z} \right) \right]^{-1}. 
\end{align}

Varying the  initial guess $\hat\mu_0$,  we end up with  few thousands
reconstructions and calculate their $\chi^2$ as
\begin{align}
\chi^2_{\text{SN},n} & = \bm{\delta\mu}_n^\mathrm{T} \tens{C}^{-1} \bm{\delta\mu}_n,
\intertext{where}
\bm{\delta\mu}_n & = \hat{\bm{\mu}}_n - \bm{\mu}_i
\end{align}
is the residual vector for a given reconstruction $n$ and \tens{C} is the covariance matrix provided by \citet{2014A&A...568A..22B}. 
We then only keep reconstructions such that $\chi^2_\text{SN}<\chi^2_\mathrm{SN,\Lambda CDM}$.  
These reconstructions  represent a non-exhaustive sample  of plausible expansion histories. 

We should note that the method of smoothing we used in this work is in fact insensitive to the initial conditions and choice of the smoothing scale \citep[c.f.][]{2006MNRAS.366.1081S, 2007MNRAS.380.1573S}: whatever the initial conditions, the method converges to the solution preferred by the data. However, they will approach this final solution via different paths. The central idea of using the iterative smoothing in this work  is to come with a \emph{non-exhaustive} sample of plausible expansion histories of the Universe directly reconstructed from the data, therefore we start the procedure with  several initial conditions and combine the results at the end.

\begin{figure}
  \centering
  \includegraphics[width=\columnwidth]{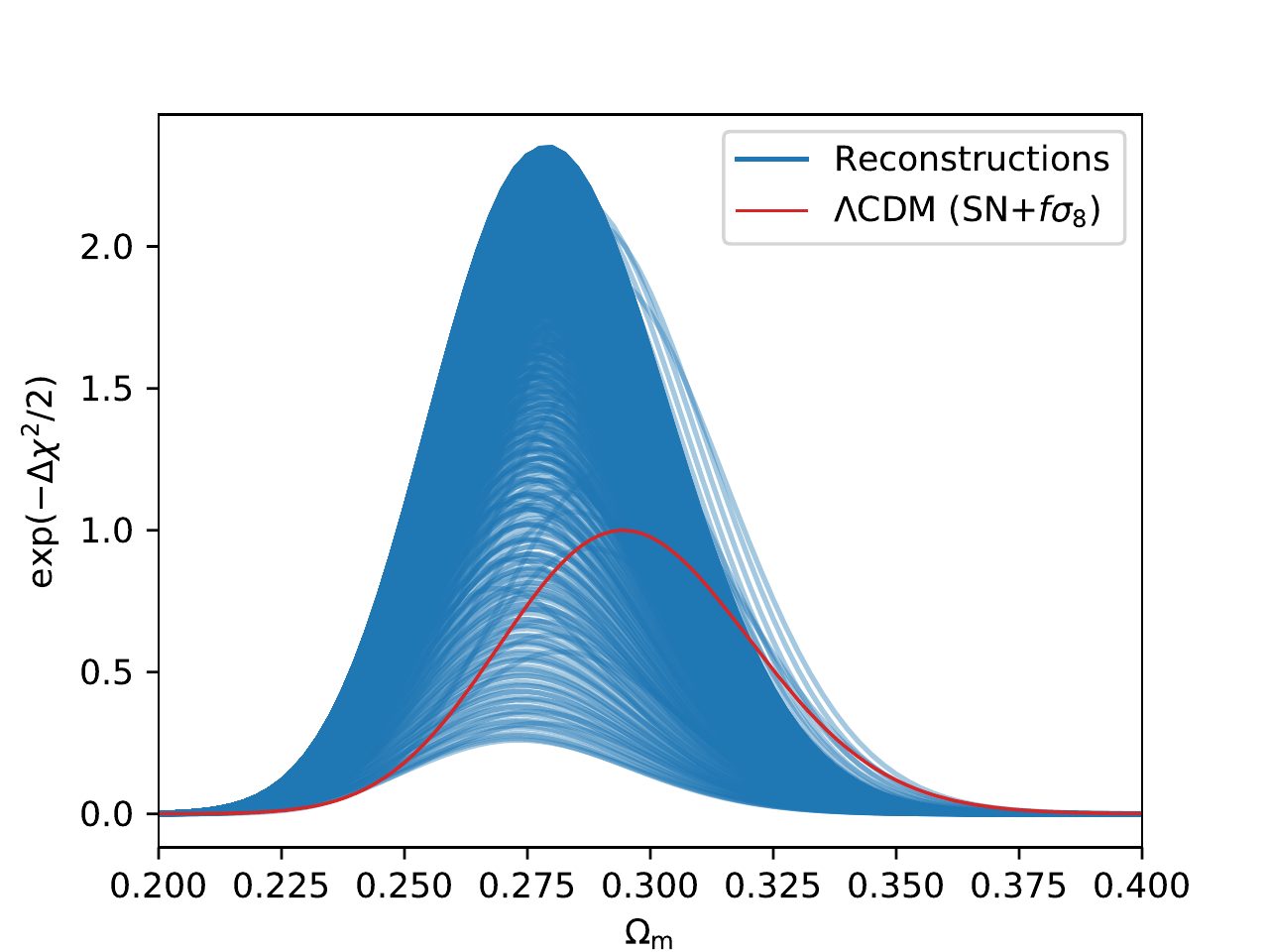}
  \caption{\label{fig:pOm}$\exp(-\Delta\chi^2/2)$  (with respect to the best-fit \lcdm\ model) versus   \Omm\   for   
  each reconstruction, fixing $(\gamma,\sigma_8) = (0.55,0.80)$.
  The red line shows the \lcdm\ case.}
\end{figure}

\subsection{Combining the likelihoods}
\label{sec:chi2}

For each reconstructed $h_n(z)$, we can calculate \fseight\ for some $(\Omm,\gamma,\sigma_8)$ 
by computing the integral in eq.~\eqref{eq:fs8}. 
We can thus explore the parameter space, and compare to 
growth measurements to obtain a \chsq\ for the growth data.
We used a compilation of growth data 
points from 
2dFGRS \citep{2009JCAP...10..004S},
WiggleZ \citep{2011MNRAS.415.2876B},
6dFGRS \citep{2012MNRAS.423.3430B},
the VIPERS \citep{2013MNRAS.435..743D},
the SDSS Main galaxy sample \citep{2015MNRAS.449..848H}, 
2MTF \citep{2017MNRAS.471.3135H},
and BOSS DR12 \citep{2017MNRAS.465.1757G}.
We did  not include the FastSound  data \citep{2016PASJ...68...38O} at
$z=1.4$, since our smooth reconstructions do not reach that redshift.

Since both datasets are independent,  we can multiply the likelihood,
or equivalently sum the $\chi^2$.
Since  the growth  data are  mutually independent,  their $\chi^2$  is
simply defined as 
\begin{align}
  \chi^2_{\fseight} & =
  \sum_i{\left(
    \frac{\hat{\fseight}(z_i | \gamma, \Omm, \sigma_8) - f\sigma_{8,i}} 
         {\sigma_{\fseight,i}}
         \right)^2}.
\end{align}

The total $\chi^2_n$ for reconstruction $n$ is thus
$\chi^2_n = \chi^2_{\mathrm{SN},n} + \chi^2_{\fseight,n}$. 
We can then find the parameters  that minimize the $\chi^2$, and their
associated confidence intervals.

\begin{figure*}
  \centering
  \includegraphics[width=\textwidth]{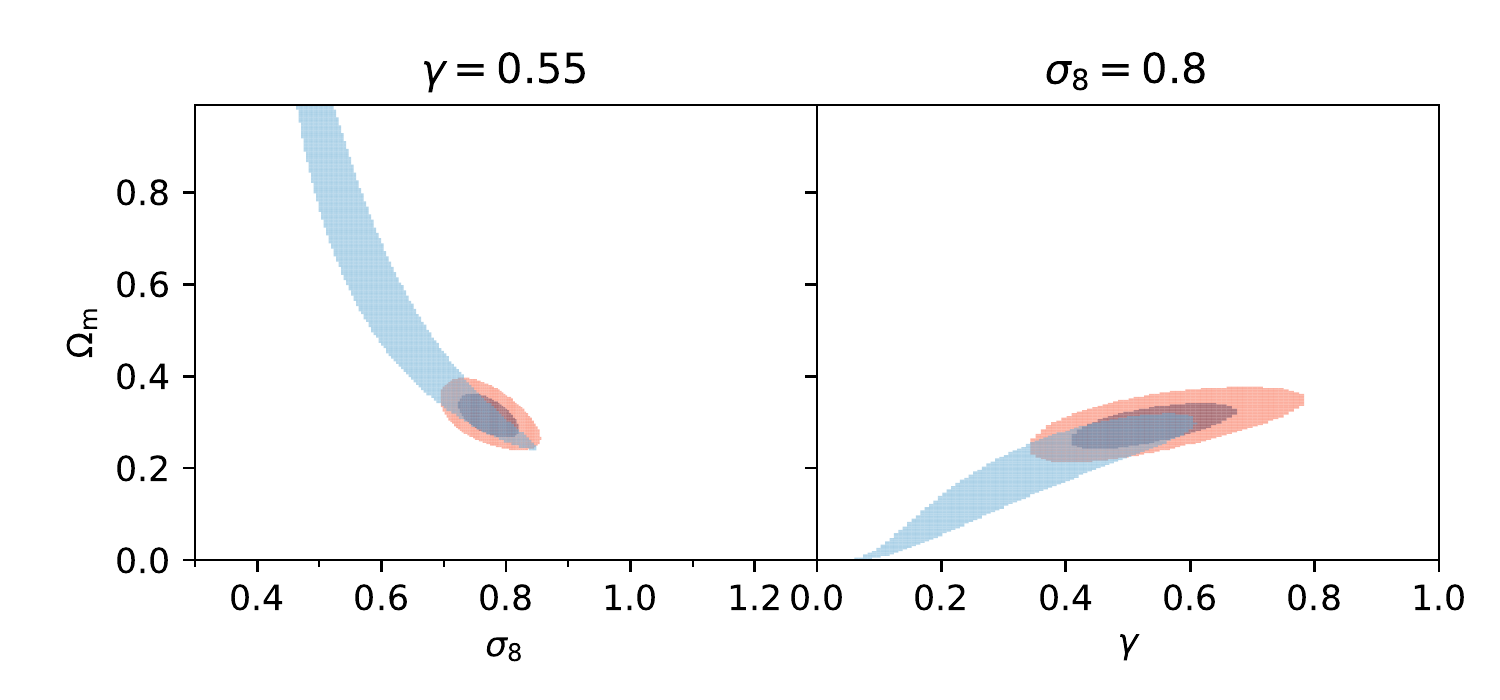}
  \caption{\label{fig:2param} Superposition of the $\Delta\chi^2<0$  
  (with respect to the best-fit \lcdm\ model) regions for 
    $\gamma=  0.55$   (left)  and   $\sigma_8=0.80$  (right)   in  the
    model-independent case in blue.
    We also  show in red  the $1\sigma$  and $2\sigma$ regions  of the
    \lcdm\ model.
  }
\end{figure*}

\section{Results}
\label{sec:res}

Using the  reconstructed expansion histories $h(z)$,  we calculate the
$\chi^2$ as defined in \S~\ref {sec:chi2}.
First, we fixed $(\gamma,\sigma_8) =  (0.55,0.80)$, and allow \Omm\ to
vary.
Since the reconstructed $h(z)$ were obtained assuming a flat universe,
\Omm\ is allowed to vary between 0 and 1.
For reference, we calculate the $\chi^2$ of the \lcdm\ model, and find 
its minimum $\chi^2_\text{min,\lcdm}$. 
We are interested in $\Delta\chi^2 = \chi^2-\chi^2_\text{min,\lcdm}$,  
the difference with respect to the best-fit \lcdm\ case. 
Fig.~\ref{fig:pOm} shows   $\mathcal{L} = \exp(-\Delta\chi^2/2)$  as a
function   of  \Omm\   for  each   reconstruction  (in   blue).
Therefore, combinations of $h$ and $\Omm$ with
a better $\chi^2$ than the best-fit \lcdm\ model ($\Delta\chi^2<0$),  have a likelihood larger than one. 
For  comparison,   we  also   show  in red $\mathcal{L}_\text{\lcdm} = \exp(-\Delta\chi^2/2)$   for  the
\lcdm\ case.    
The  model-independent  reconstructions  seem to  favour  slightly  lower
\Omm\ with respect to the \lcdm\ case. 
However, they are fully consistent with the \lcdm\ case.

We then allow $\gamma$ or $\sigma_8$ to vary together with \Omm, while
fixing the third  parameter to its fiducial  value ($\sigma_8=0.80$ or
$\gamma = 0.55$). 
In both cases, we calculate $\chi^2$ for the \lcdm\ case, and find the regions where $\Delta\chi^2< 2.3$ and  $\Delta\chi^2< 6.18$, corresponding to $1\sigma$ and $2\sigma$ for two degrees of freedom. 
Fig.~\ref{fig:2param} shows in red the $1\sigma$ and $2\sigma$ regions of the \lcdm\ case. 
For each model-independent reconstruction, we then calculate the $\chi^2$ of the model-independent case, and find the regions in   the  $(\sigma_8,\Omm)$   and   $(\gamma,\Omm)$
planes where the reconstruction give a better $\chi^2$ than the best-fit \lcdm, namely, $\Delta\chi^2<0$.
Fig.~\ref{fig:2param} shows  in blue the superposition of these regions over all
reconstructions in the $(\Omm,\sigma_8)$ (left) and $(\Omm,\gamma)$
(right) planes. 
Therefore, if a  point $(\sigma_8,\Omm)$ (or $(\gamma,\Omm)$) is
located in the blue region,  there exists at least
one reconstruction that, combined with $(\sigma_8,\Omm)$  (or
$(\gamma,\Omm)$), yields a better $\chi^2$ than the best-fit
\lcdm\ model.

\begin{figure*}
  \centering
  \includegraphics[width=\textwidth]{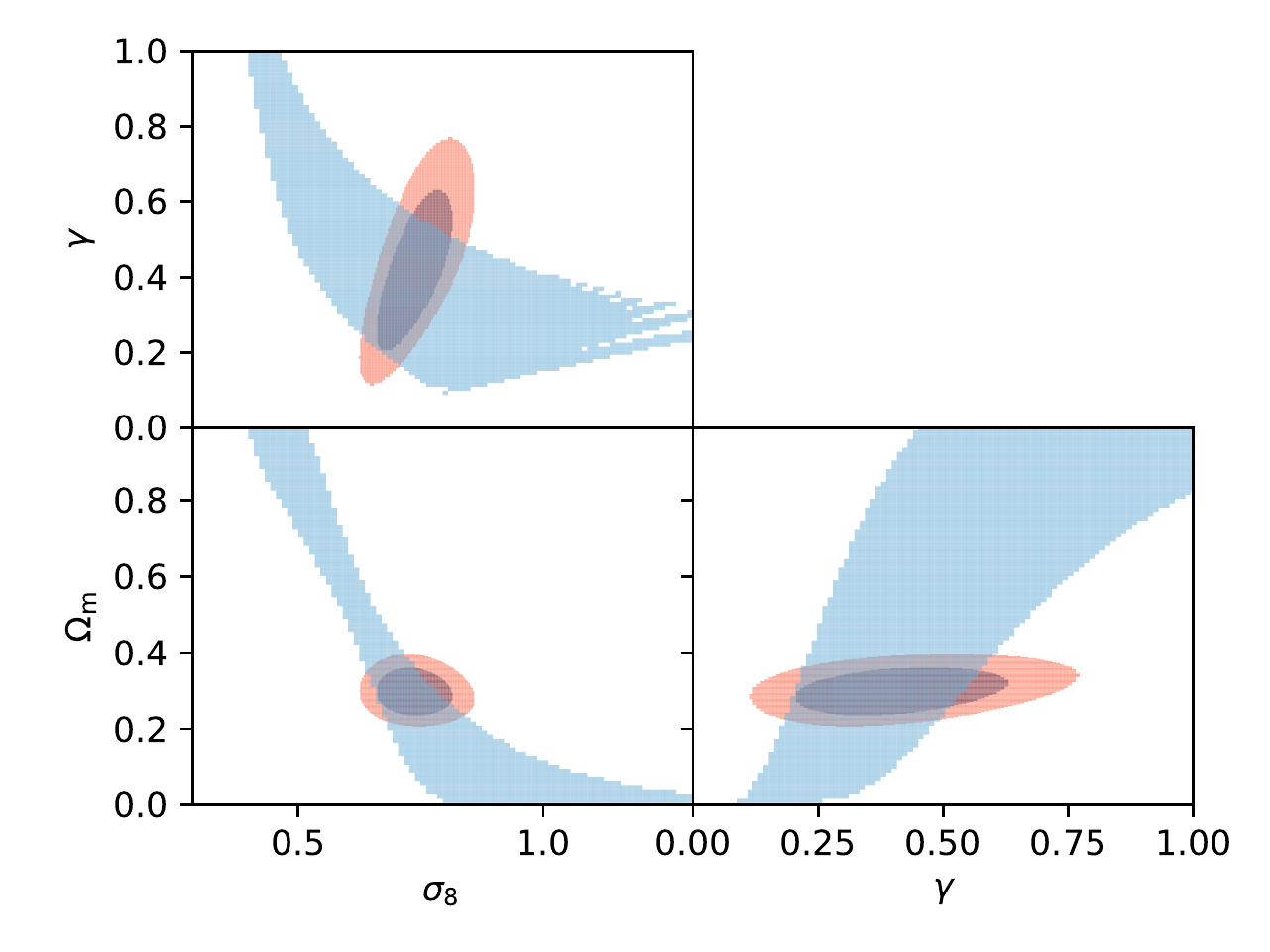}
  \caption{\label{fig:3param}Superposition of the $\Delta\chi^2<0$  (with respect to the best-fit \lcdm\ model) regions     
  for   $(\Omm,\gamma,\sigma_8)$    for   the  model-independent case (blue). 
  In red we show the $1\sigma$ and $2\sigma$ regions for the \lcdm\ case.
  } 
\end{figure*}

Fixing $\gamma=0.55$ yields higher preferred values for \Omm,
while fixing $\sigma_8=0.80$ yields lower preferred values.
However, the model-independent approach is fully consistent with \lcdm.  
Moreover, it can be seen that, when $h(z)$ is not restricted to \lcdm, there is
a stronger degeneracy in the parameters.
Namely, it is possible to find expansions histories that, coupled with
low values  of \Omm\ and $\gamma$,  or with high values  of $\Omm$ and
$\sigma_8$, give a better  fit to the combined data. 
The degeneracy in the parameters can be understood from eq.~\ref{eq:fs8}: for fixed $\sigma_8$, 
$\Omm^\gamma$ should stay roughly constant, , therefore lower $\Omm$ are 
compensated by lower $\gamma$.
Similarly,  for fixed $\gamma$, $\Omm\sigma_8$ should stay constant, therefore
lower $\Omm$ demand higher higher $\sigma_8$.
This is  consistent with  the results  of \citet{2013PhRvD..87b3520S},
with slightly tighter constraints.

Finally, we vary all three parameters $(\Omm,\gamma,\sigma_8)$ simultaneously.
Fig.~\ref{fig:3param} shows in red the projections of the  $\Delta\chi^2<3.53$ and 8.02 regions of the \lcdm\ case,  
corresponding to $1\sigma$ and $2\sigma$ for three degrees of freedom, onto
the  $(\sigma_8,\gamma)$ (top-left),  $(\sigma_8,\Omm)$ (bottom-left),
and $(\gamma,\Omm)$ (bottom-right).
For the model-independent case, we proceed as in Fig.~\ref{fig:2param}, and find the $\Delta\chi^2<0$ regions for each reconstruction. 
We then show in blue the projection onto the three planes of the superposition of the $\Delta\chi^2<0$ regions over all reconstruction. 
Again,   the blue region shows the region of the parameter-space where there is at least one model-independent reconstruction that yields a better $\chi^2$ than the best-fit \lcdm\ model. 

The  model-independent joint  constraints on  $(\Omm,\gamma,\sigma_8)$
are now very  broad.
They are fully consistent with the \lcdm\ model.
The $\Delta\chi^2<0$ region is consistent with both $\Omm=0$ and
$\Omm =  1$, while it allows $\gamma$    between about 0.1 and  1, and $\sigma_8$
 between 0.25 and 1.25.

\section{Dark energy constraints}
\label{sec:pos}

\begin{figure}
  \centering
  \includegraphics[width=\columnwidth]{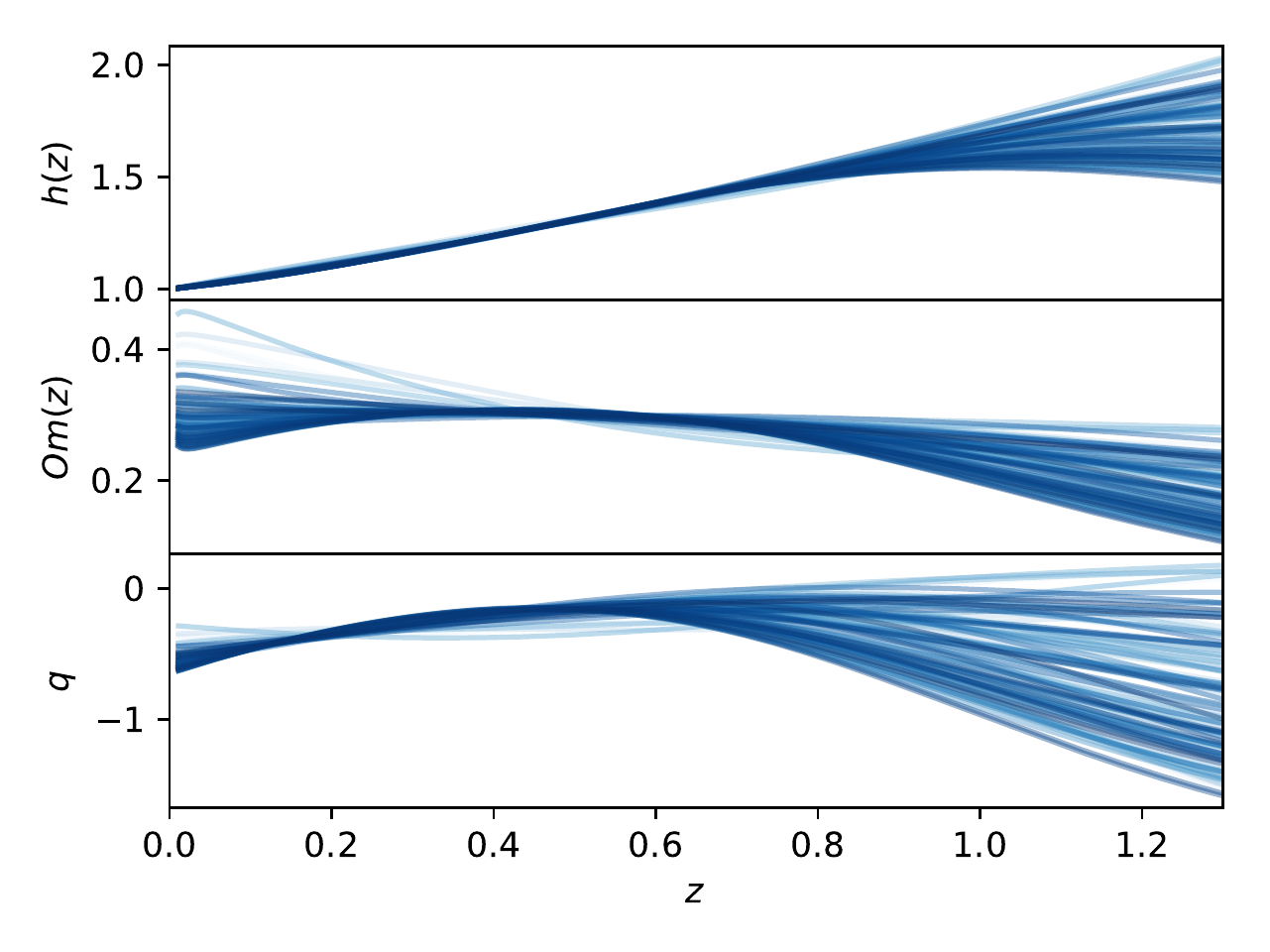}
  \caption{\label{fig:qhOm}%
    Reconstructed $h(z)$, $Om(z)$, and $q(z)$.
    The colour-code shows the index of the reconstruction.}
\end{figure}

\begin{figure}
  \centering
  \includegraphics[width=\columnwidth]{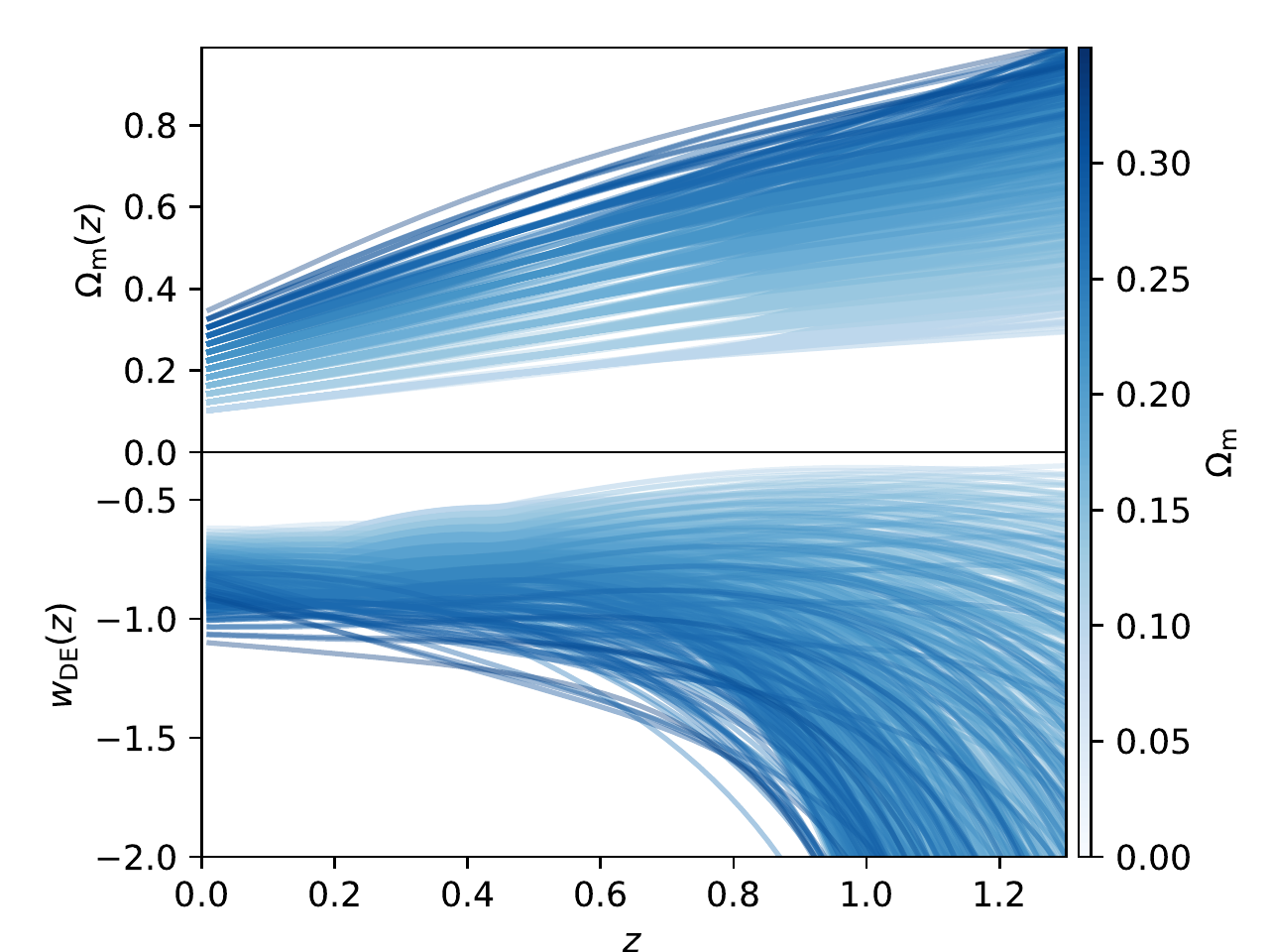}
  \caption{\label{fig:wde}%
    Reconstructed $\Omm(z)$ (top) and $w(z)$ (bottom)  for
    different \Omm\ and $h(z)$. 
  All lines here verify eq.~\eqref{eq:valid}.
    }
\end{figure}

\begin{figure*}
  \centering
  \includegraphics[width=\textwidth]{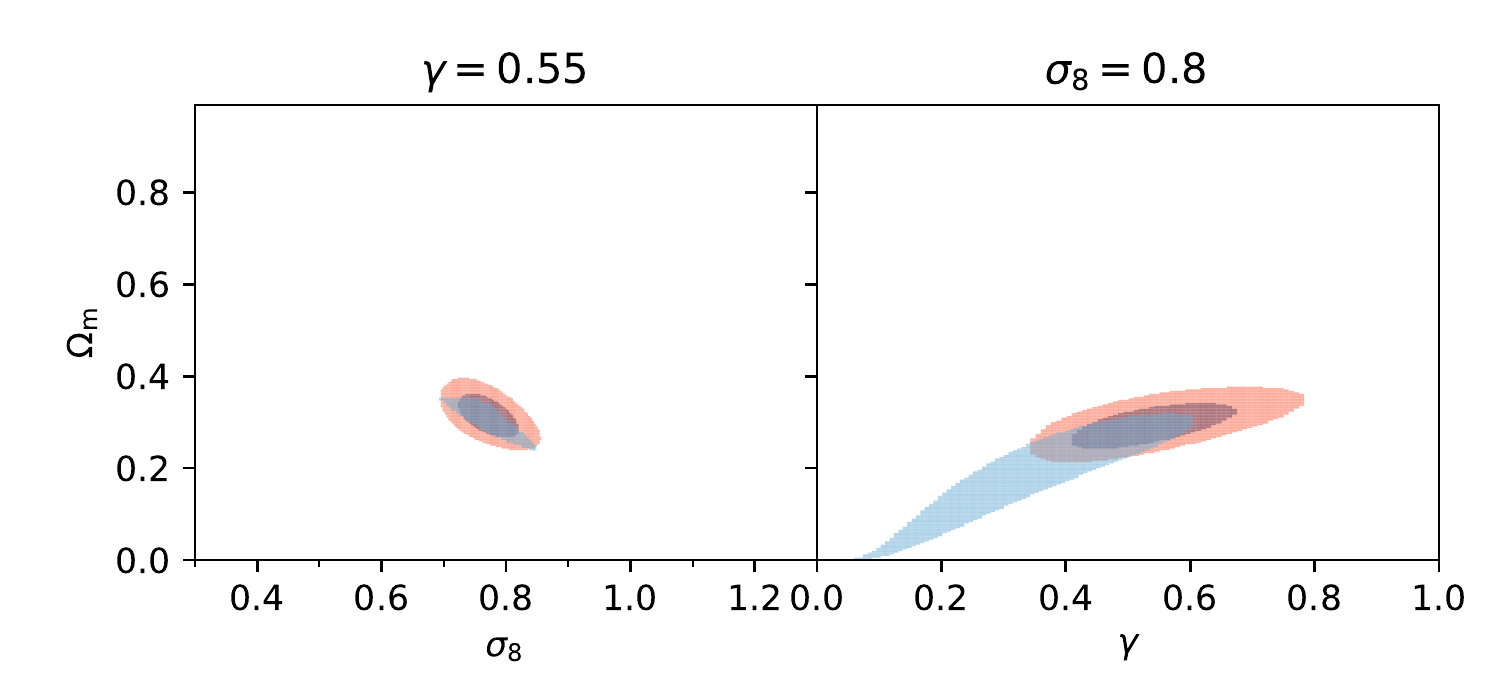}
  \caption{\label{fig:2param_wde} Blue: 
 Truncation of the $\Delta\chi^2<0$  (with respect to the best-fit \lcdm\ model) regions for 
    $\gamma=  0.55$   (left)  and   $\sigma_8=0.80$  (right)   in  the
    model-independent case using eq.~\eqref{eq:valid}  as a
    hard prior.
    Red: $1\sigma$  and $2\sigma$ regions  of the \lcdm\ model.    
  }
\end{figure*}


In   the  previous   section,  we   considered  all   combinations  of
$(\Omm,h(z))$, with  the only  restriction $\Omm<1$, since  the $h(z)$
were obtained assuming a flat universe.
Rewriting equation~\eqref{eq:h} as
\begin{equation}
  \Omega_\mathrm{de}(z) = h^2(z)-\Omm(1+z)^3,
  \end{equation}
another  constraint arises.
Namely, the equation of state $w$ is well defined, i.e., does not have a singularity, if $\Omde(z)>0$ at all redshift.

Therefore, even though some models can have negative DE density
\citep[e.g.]{2003JCAP...11..014S,2014ApJ...793L..40S},  
in  this section, we only consider  combinations  of   $h(z)$  and
$\Omm$  respecting  the positivity condition  
\begin{equation}
  \label{eq:valid}
  h^2(z)-\Omm\,(1+z)^3\geq 0
\end{equation}
for all $z$.

We can then use this to reconstruct the dark energy equation of state
\begin{align}
  \begin{split}
  w_\mathrm{de}(z)& = \frac{ \frac 2 3 (1+z) \frac{h'(z)}{h(z)}-1}
  {1-\Omm(1+z)^3 h^{-2}}\\
  & = \frac 1 3 \frac{2q-1}{1-\Omm(z)},
  \end{split}
  \intertext{where}
  q(z) &= (1+z)\frac{h'(z)}{h(z)}-1
\end{align}
is the deceleration parameter.

The top-, middle-, and bottom panels  of Fig.~\ref{fig:qhOm} show the
expansion history $h(z)$,  the \Om\ parameter
\citep{2008PhRvD..78j3502S} 
\begin{align}
\Om(z) &= \frac{h^2(z)-1}{(1+z)^3-1},
\end{align}
and the  deceleration parameter $q(z)$ for a random choice of about
5\% of the  reconstructions. 
For a flat \lcdm\ Universe, $\Om(z) \equiv \Omm$, thus \Om\ is a litmus
test for flat \lcdm.  

The top and  bottom  panels  of Fig.~\ref{fig:wde} respectively  show
the  matter  density $\Omm(z)$  and  the  equation  of   state  of
DE  for  some combination of $(\Omm,h(z))$ verifying the
positivity condition \eqref{eq:valid}, colour-coded by the index of the
reconstruction.    
The expansion histories that are closer from \lcdm\ have a
 \Om\ parameter close to constant, and their $q(z)$ can cross 0, while
 some reconstructions further from \lcdm\ do not cross 0.  
When $\Omm(z)$ crosses 1, eq.~\eqref{eq:valid} ceases to be valid,
therefore none of the lines shown here crosses 1. 

\begin{figure*}
  \centering
  \includegraphics[width=\textwidth]{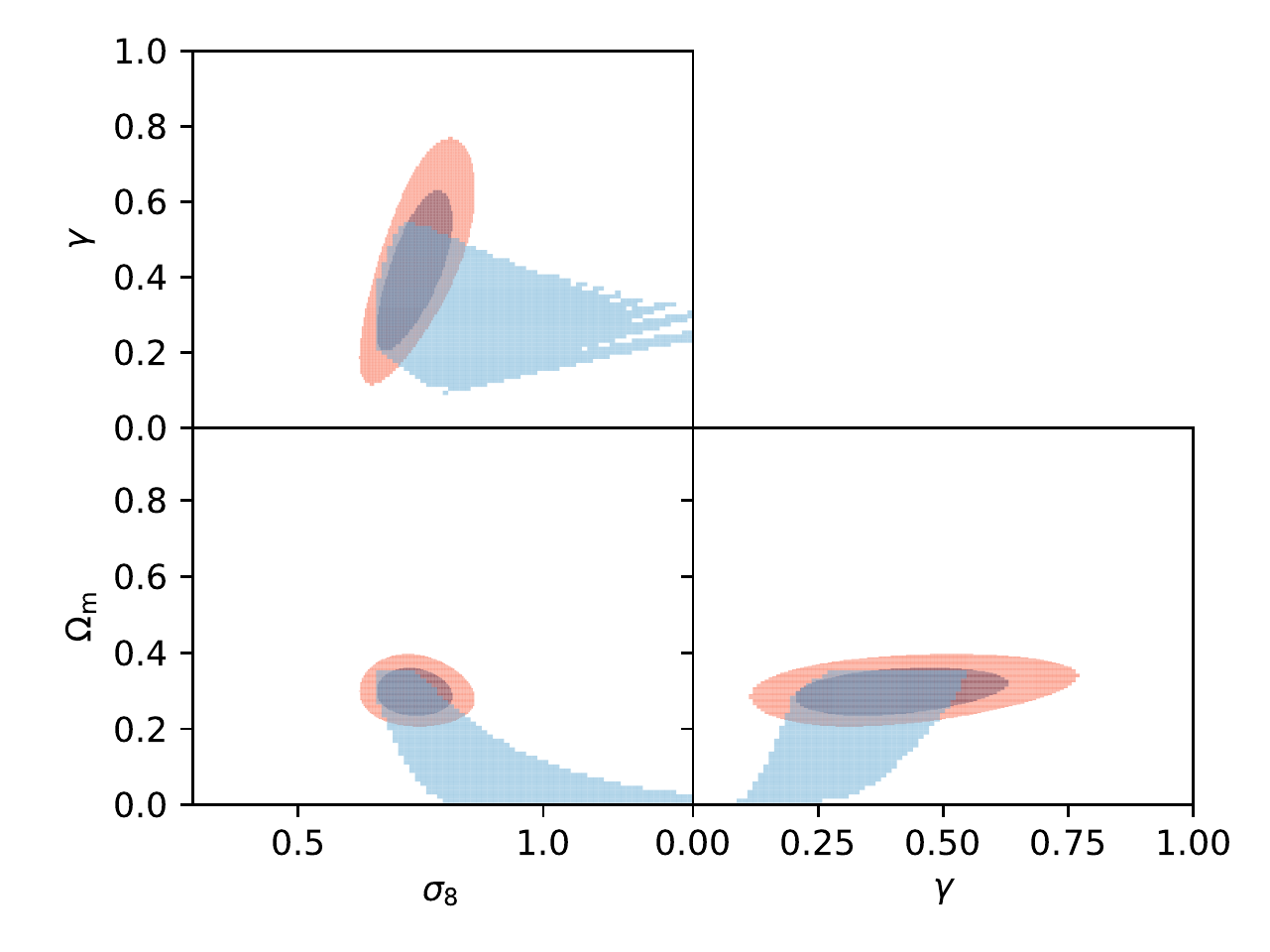}
  \caption{\label{fig:3param_wde}
    Blue: Truncation   of    the   $\Delta\chi^2<0$   (with respect to the best-fit \lcdm\ model) regions    for   $(\Omm,\gamma,\sigma_8)$    for   the
    model-independent case  using eq.~\eqref{eq:valid}  as a
    hard prior.
    Red: $1\sigma$ and $2\sigma$ regions of the \lcdm\ case.
  }
\end{figure*}

We can now add the positivity condition \eqref{eq:valid}  as a hard
prior on \Omm\ in the previous analysis.
Indeed, large values  of \Omm\ combined with  some reconstructions can
lead  to  negative  DE density,  and   these  combinations  should  thus  be
rejected. 
Figs.~\ref{fig:2param_wde} and  \ref{fig:3param_wde} show in  blue the
superposition over all reconstructions verifying
equation~\eqref{eq:valid} of the projected  $\Delta\chi^2<0$
regions of  the parameter  space.
The red contours are unchanged with respect to Figs~\ref{fig:2param} and~\ref{fig:3param}.

In  Fig.~\ref{fig:2param_wde}, while  the $\sigma_8=0.8$  case
(right-hand panel) is  not affected much,  since it preferred  lower
values of \Omm,  the  allowed region for the  $\gamma=0.55$ case is
drastically reduced,  and only a small space of the original $\Delta\chi^2<0$ 
regions (that is, before applying eq.~\eqref{eq:valid}) is allowed. 
This region is located in the $2\sigma$ region of the \lcdm\ case.

In Fig.~\ref{fig:3param_wde},  the $\Delta\chi^2<0$ regions    in   each    projection    are also  truncated    with   respect    to
Fig.~\ref{fig:3param}, restricting the lower range of $\sigma_8$ and the higher range of $\gamma$ and $\Omm$.

The positivity condition~\eqref{eq:valid}   is thus  a very  strong
constraint  on the cosmological parameters, since it forbids large
values of $\Omm\ \gtrsim 0.4$.
Indeed,  for these values, the DE density crosses zero within  
our data range, therefore these values are not allowed here.
On the other hand, for low enough values, $\Omde(z)$ never crosses zero.


\section{Discussion and conclusion}

\label{sec:ccl}

Using model-independent  reconstructions of the expansion  history from type Ia supernovae data, 
we    fit    the   growth    data    and    obtain   constraints    on
$(\Omm,\gamma,\sigma_8)$. 
These model-independent constraints on the cosmological parameters are
broader than the \lcdm\ ones, but fully consistent.
When all three cosmological parameter are  let free, they are not well
constrained,  and it  is  possible to  find  expansion histories  with
cosmological parameters  that are far  from the \lcdm\  constraints that
give a reasonable fit to the data.

However,  when   restricting  the  combinations  of   $\Omm$  and  the
reconstructed expansion  histories $h(z)$  that yield a  positive dark
energy density parameter  ($h^2(z)-\Omm(1+z)^3>0$), the constraints on
the cosmological parameters become stronger. 
Moreover,   when  imposing   GR,  i.e.,   fixing  $\gamma=0.55$,   the
model-independent contours are truncated  and fully contained within the  \lcdm\ ones,
showing strong evidence in favour of \lcdm.
That is, combinations of large \Omm\ with expansion histories that are
too different from \lcdm\ are excluded.  
It should be noted that in \citet{2005PhRvD..72d3529L}, $\gamma$ depends 
on $w$ following $\gamma(w) = 0.55 + 0.05 w(z=1)$, therefore fixing $\gamma=0.55$ 
is not completely model-independent. 
However, we expect it to have little influence on the results presented here.

Our  constraints are  more stringent  than \citet{2013PhRvD..87b3520S}
thanks to the 
better quality of the data and 
the introduction of the DE density positivity condition. 
The results are consistent with a flat-\lcdm\ Universe and gravity described by 
general relativity, although modified theories of gravity predicting different 
growth index cannot be ruled out at this stage. 
The  combined  $\chi^2$  being currently  dominated  by  the
supernovae  data, better  growth  measurements are  needed to  further
constrain gravity theory.  


Future  surveys such  as the Dark Energy Spectroscopic Instrument
\citep{2016arXiv161100036D} will  bring down   the   errors  on   the
growth   measurements,  while   surveys such as the {Wide Field Infrared Survey Telescope} 
\citep[WFIRST,][]{2015arXiv150303757S}  and the Large Synoptic Survey
Telescope  \citep[LSST,][]{2008arXiv0805.2366I} are expected to
observe thousands of supernovae, increasing the quality of the data
and covering a larger redshift range.   

\section*{Acknowledgements}
We thank  Eric Linder for useful discussions, and 
Teppei Okumura 
for providing us with the \fseight\ data. 
The  computations  were  performed   by  using  the  high  performance
computing cluster  Polaris at  the Korea  Astronomy and  Space Science
Institute.  
A.S.  would like  to acknowledge the support of  the National Research
Foundation of Korea (NRF-2016R1C1B2016478). 
The authors thank the Yukawa Institute for Theoretical Physics at Kyoto University. 
Discussions during the CosKASI-ICG-NAOC-YITP joint workshop YITP-T-17-03 were useful 
to complete this work.




\bibliographystyle{mnras}
\bibliography{biblio} 

\begin{thebibliography}{}
\makeatletter
\relax
\def\mn@urlcharsother{\let\do\@makeother \do\$\do\&\do\#\do\^\do\_\do\%\do\~}
\def\mn@doi{\begingroup\mn@urlcharsother \@ifnextchar [ {\mn@doi@}
  {\mn@doi@[]}}
\def\mn@doi@[#1]#2{\def\@tempa{#1}\ifx\@tempa\@empty \href
  {http://dx.doi.org/#2} {doi:#2}\else \href {http://dx.doi.org/#2} {#1}\fi
  \endgroup}
\def\mn@eprint#1#2{\mn@eprint@#1:#2::\@nil}
\def\mn@eprint@arXiv#1{\href {http://arxiv.org/abs/#1} {{\tt arXiv:#1}}}
\def\mn@eprint@dblp#1{\href {http://dblp.uni-trier.de/rec/bibtex/#1.xml}
  {dblp:#1}}
\def\mn@eprint@#1:#2:#3:#4\@nil{\def\@tempa {#1}\def\@tempb {#2}\def\@tempc
  {#3}\ifx \@tempc \@empty \let \@tempc \@tempb \let \@tempb \@tempa \fi \ifx
  \@tempb \@empty \def\@tempb {arXiv}\fi \@ifundefined
  {mn@eprint@\@tempb}{\@tempb:\@tempc}{\expandafter \expandafter \csname
  mn@eprint@\@tempb\endcsname \expandafter{\@tempc}}}

\bibitem[\protect\citeauthoryear{{Alam} et~al.,}{{Alam}
  et~al.}{2017}]{2017MNRAS.470.2617A}
{Alam} S.,  et~al., 2017, \mn@doi [\mnras] {10.1093/mnras/stx721}, \href
  {http://adsabs.harvard.edu/abs/2017MNRAS.470.2617A} {470, 2617}

\bibitem[\protect\citeauthoryear{{Basilakos}}{{Basilakos}}{2012}]{2012IJMPD..2150064B}
{Basilakos} S.,  2012, \mn@doi [International Journal of Modern Physics D]
  {10.1142/S0218271812500642}, \href
  {http://adsabs.harvard.edu/abs/2012IJMPD..2150064B} {21, 1250064}

\bibitem[\protect\citeauthoryear{{Bean} \& {Tangmatitham}}{{Bean} \&
  {Tangmatitham}}{2010}]{2010PhRvD..81h3534B}
{Bean} R.,  {Tangmatitham} M.,  2010, \mn@doi [\prd]
  {10.1103/PhysRevD.81.083534}, \href
  {http://adsabs.harvard.edu/abs/2010PhRvD..81h3534B} {81, 083534}

\bibitem[\protect\citeauthoryear{{Bennett} et~al.,}{{Bennett}
  et~al.}{2003}]{2003ApJS..148....1B}
{Bennett} C.~L.,  et~al., 2003, \mn@doi [\apjs] {10.1086/377253}, \href
  {http://adsabs.harvard.edu/abs/2003ApJS..148....1B} {148, 1}

\bibitem[\protect\citeauthoryear{{Betoule} et~al.,}{{Betoule}
  et~al.}{2014}]{2014A&A...568A..22B}
{Betoule} M.,  et~al., 2014, \mn@doi [\aap] {10.1051/0004-6361/201423413},
  \href {http://adsabs.harvard.edu/abs/2014A%26A...568A..22B} {568, A22}

\bibitem[\protect\citeauthoryear{{Beutler} et~al.,}{{Beutler}
  et~al.}{2012}]{2012MNRAS.423.3430B}
{Beutler} F.,  et~al., 2012, \mn@doi [\mnras]
  {10.1111/j.1365-2966.2012.21136.x}, \href
  {http://adsabs.harvard.edu/abs/2012MNRAS.423.3430B} {423, 3430}

\bibitem[\protect\citeauthoryear{{Blake} et~al.,}{{Blake}
  et~al.}{2011}]{2011MNRAS.415.2876B}
{Blake} C.,  et~al., 2011, \mn@doi [\mnras] {10.1111/j.1365-2966.2011.18903.x},
  \href {http://adsabs.harvard.edu/abs/2011MNRAS.415.2876B} {415, 2876}

\bibitem[\protect\citeauthoryear{{DESI Collaboration} et~al.,}{{DESI
  Collaboration} et~al.}{2016}]{2016arXiv161100036D}
{DESI Collaboration} et~al., 2016, preprint, \href
  {http://adsabs.harvard.edu/abs/2016arXiv161100036D} {} (\mn@eprint {arXiv}
  {1611.00036})

\bibitem[\protect\citeauthoryear{{Dvali}, {Gabadadze}  \& {Porrati}}{{Dvali}
  et~al.}{2000}]{2000PhLB..485..208D}
{Dvali} G.,  {Gabadadze} G.,   {Porrati} M.,  2000, \mn@doi [Physics Letters B]
  {10.1016/S0370-2693(00)00669-9}, \href
  {http://adsabs.harvard.edu/abs/2000PhLB..485..208D} {485, 208}

\bibitem[\protect\citeauthoryear{{Eisenstein} et~al.,}{{Eisenstein}
  et~al.}{2005}]{2005ApJ...633..560E}
{Eisenstein} D.~J.,  et~al., 2005, \mn@doi [\apj] {10.1086/466512}, \href
  {http://adsabs.harvard.edu/abs/2005ApJ...633..560E} {633, 560}

\bibitem[\protect\citeauthoryear{{Gil-Mar{\'{\i}}n}, {Percival}, {Verde},
  {Brownstein}, {Chuang}, {Kitaura}, {Rodr{\'{\i}}guez-Torres}  \&
  {Olmstead}}{{Gil-Mar{\'{\i}}n} et~al.}{2017}]{2017MNRAS.465.1757G}
{Gil-Mar{\'{\i}}n} H.,  {Percival} W.~J.,  {Verde} L.,  {Brownstein} J.~R.,
  {Chuang} C.-H.,  {Kitaura} F.-S.,  {Rodr{\'{\i}}guez-Torres} S.~A.,
  {Olmstead} M.~D.,  2017, \mn@doi [\mnras] {10.1093/mnras/stw2679}, \href
  {http://adsabs.harvard.edu/abs/2017MNRAS.465.1757G} {465, 1757}

\bibitem[\protect\citeauthoryear{{G{\'o}mez-Valent}, {Sol{\`a}}  \&
  {Basilakos}}{{G{\'o}mez-Valent} et~al.}{2015}]{2015JCAP...01..004G}
{G{\'o}mez-Valent} A.,  {Sol{\`a}} J.,   {Basilakos} S.,  2015, \mn@doi [\jcap]
  {10.1088/1475-7516/2015/01/004}, \href
  {http://adsabs.harvard.edu/abs/2015JCAP...01..004G} {1, 004}

\bibitem[\protect\citeauthoryear{{Howlett}, {Ross}, {Samushia}, {Percival}  \&
  {Manera}}{{Howlett} et~al.}{2015}]{2015MNRAS.449..848H}
{Howlett} C.,  {Ross} A.~J.,  {Samushia} L.,  {Percival} W.~J.,   {Manera} M.,
  2015, \mn@doi [\mnras] {10.1093/mnras/stu2693}, \href
  {http://adsabs.harvard.edu/abs/2015MNRAS.449..848H} {449, 848}

\bibitem[\protect\citeauthoryear{{Howlett} et~al.,}{{Howlett}
  et~al.}{2017}]{2017MNRAS.471.3135H}
{Howlett} C.,  et~al., 2017, \mn@doi [\mnras] {10.1093/mnras/stx1521}, \href
  {http://adsabs.harvard.edu/abs/2017MNRAS.471.3135H} {471, 3135}

\bibitem[\protect\citeauthoryear{{Ivezic} et~al.,}{{Ivezic}
  et~al.}{2008}]{2008arXiv0805.2366I}
{Ivezic} Z.,  et~al., 2008, preprint, \href
  {http://adsabs.harvard.edu/abs/2008arXiv0805.2366I} {} (\mn@eprint {arXiv}
  {0805.2366})

\bibitem[\protect\citeauthoryear{{L'Huillier} \& {Shafieloo}}{{L'Huillier} \&
  {Shafieloo}}{2017}]{2017JCAP...01..015L}
{L'Huillier} B.,  {Shafieloo} A.,  2017, \mn@doi [\jcap]
  {10.1088/1475-7516/2017/01/015}, \href
  {http://adsabs.harvard.edu/abs/2017JCAP...01..015L} {1, 015}

\bibitem[\protect\citeauthoryear{{Linder}}{{Linder}}{2005}]{2005PhRvD..72d3529L}
{Linder} E.~V.,  2005, \mn@doi [\prd] {10.1103/PhysRevD.72.043529}, \href
  {http://adsabs.harvard.edu/abs/2005PhRvD..72d3529L} {72, 043529}

\bibitem[\protect\citeauthoryear{{Linder}}{{Linder}}{2017}]{2017APh....86...41L}
{Linder} E.~V.,  2017, \mn@doi [Astroparticle Physics]
  {10.1016/j.astropartphys.2016.11.002}, \href
  {http://adsabs.harvard.edu/abs/2017APh....86...41L} {86, 41}

\bibitem[\protect\citeauthoryear{{Linder} \& {Cahn}}{{Linder} \&
  {Cahn}}{2007}]{2007APh....28..481L}
{Linder} E.~V.,  {Cahn} R.~N.,  2007, \mn@doi [Astroparticle Physics]
  {10.1016/j.astropartphys.2007.09.003}, \href
  {http://adsabs.harvard.edu/abs/2007APh....28..481L} {28, 481}

\bibitem[\protect\citeauthoryear{{Mueller}, {Percival}, {Linder}, {Alam},
  {Zhao}, {S{\'a}nchez}  \& {Beutler}}{{Mueller}
  et~al.}{2016}]{2016arXiv161200812M}
{Mueller} E.-M.,  {Percival} W.,  {Linder} E.,  {Alam} S.,  {Zhao} G.-B.,
  {S{\'a}nchez} A.~G.,   {Beutler} F.,  2016, preprint, \href
  {http://adsabs.harvard.edu/abs/2016arXiv161200812M} {} (\mn@eprint {arXiv}
  {1612.00812})

\bibitem[\protect\citeauthoryear{{Nesseris} \& {Perivolaropoulos}}{{Nesseris}
  \& {Perivolaropoulos}}{2008}]{2008PhRvD..77b3504N}
{Nesseris} S.,  {Perivolaropoulos} L.,  2008, \mn@doi [\prd]
  {10.1103/PhysRevD.77.023504}, \href
  {http://adsabs.harvard.edu/abs/2008PhRvD..77b3504N} {77, 023504}

\bibitem[\protect\citeauthoryear{{Nesseris}, {Pantazis}  \&
  {Perivolaropoulos}}{{Nesseris} et~al.}{2017}]{2017PhRvD..96b3542N}
{Nesseris} S.,  {Pantazis} G.,   {Perivolaropoulos} L.,  2017, \mn@doi [\prd]
  {10.1103/PhysRevD.96.023542}, \href
  {http://adsabs.harvard.edu/abs/2017PhRvD..96b3542N} {96, 023542}

\bibitem[\protect\citeauthoryear{{Okumura} et~al.,}{{Okumura}
  et~al.}{2016}]{2016PASJ...68...38O}
{Okumura} T.,  et~al., 2016, \mn@doi [\pasj] {10.1093/pasj/psw029}, \href
  {http://adsabs.harvard.edu/abs/2016PASJ...68...38O} {68, 38}

\bibitem[\protect\citeauthoryear{{Ooba}, {Ratra}  \& {Sugiyama}}{{Ooba}
  et~al.}{2017}]{2017arXiv171003271O}
{Ooba} J.,  {Ratra} B.,   {Sugiyama} N.,  2017, preprint, \href
  {http://adsabs.harvard.edu/abs/2017arXiv171003271O} {} (\mn@eprint {arXiv}
  {1710.03271})

\bibitem[\protect\citeauthoryear{{Peebles} \& {Ratra}}{{Peebles} \&
  {Ratra}}{2003}]{2003RvMP...75..559P}
{Peebles} P.~J.,  {Ratra} B.,  2003, \mn@doi [Reviews of Modern Physics]
  {10.1103/RevModPhys.75.559}, \href
  {http://adsabs.harvard.edu/abs/2003RvMP...75..559P} {75, 559}

\bibitem[\protect\citeauthoryear{{Perlmutter} et~al.,}{{Perlmutter}
  et~al.}{1999}]{1999ApJ...517..565P}
{Perlmutter} S.,  et~al., 1999, \mn@doi [\apj] {10.1086/307221}, \href
  {http://adsabs.harvard.edu/abs/1999ApJ...517..565P} {517, 565}

\bibitem[\protect\citeauthoryear{{Planck Collaboration XIII}}{{Planck
  Collaboration XIII}}{2016}]{2016A&A...594A..13P}
{Planck Collaboration XIII} 2016, \mn@doi [\aap] {10.1051/0004-6361/201525830},
  \href {http://adsabs.harvard.edu/abs/2016A%26A...594A..13P} {594, A13}

\bibitem[\protect\citeauthoryear{{Riess} et~al.,}{{Riess}
  et~al.}{1998}]{1998AJ....116.1009R}
{Riess} A.~G.,  et~al., 1998, \mn@doi [\aj] {10.1086/300499}, \href
  {http://adsabs.harvard.edu/abs/1998AJ....116.1009R} {116, 1009}

\bibitem[\protect\citeauthoryear{{Ruiz} \& {Huterer}}{{Ruiz} \&
  {Huterer}}{2015}]{2015PhRvD..91f3009R}
{Ruiz} E.~J.,  {Huterer} D.,  2015, \mn@doi [\prd]
  {10.1103/PhysRevD.91.063009}, \href
  {http://adsabs.harvard.edu/abs/2015PhRvD..91f3009R} {91, 063009}

\bibitem[\protect\citeauthoryear{{Sahni} \& {Shtanov}}{{Sahni} \&
  {Shtanov}}{2003}]{2003JCAP...11..014S}
{Sahni} V.,  {Shtanov} Y.,  2003, \mn@doi [\jcap]
  {10.1088/1475-7516/2003/11/014}, \href
  {http://adsabs.harvard.edu/abs/2003JCAP...11..014S} {11, 014}

\bibitem[\protect\citeauthoryear{{Sahni}, {Shafieloo}  \&
  {Starobinsky}}{{Sahni} et~al.}{2008}]{2008PhRvD..78j3502S}
{Sahni} V.,  {Shafieloo} A.,   {Starobinsky} A.~A.,  2008, \mn@doi [\prd]
  {10.1103/PhysRevD.78.103502}, \href
  {http://adsabs.harvard.edu/abs/2008PhRvD..78j3502S} {78, 103502}

\bibitem[\protect\citeauthoryear{{Sahni}, {Shafieloo}  \&
  {Starobinsky}}{{Sahni} et~al.}{2014}]{2014ApJ...793L..40S}
{Sahni} V.,  {Shafieloo} A.,   {Starobinsky} A.~A.,  2014, \mn@doi [\apjl]
  {10.1088/2041-8205/793/2/L40}, \href
  {http://adsabs.harvard.edu/abs/2014ApJ...793L..40S} {793, L40}

\bibitem[\protect\citeauthoryear{{Shafieloo}}{{Shafieloo}}{2007}]{2007MNRAS.380.1573S}
{Shafieloo} A.,  2007, \mn@doi [\mnras] {10.1111/j.1365-2966.2007.12175.x},
  \href {http://adsabs.harvard.edu/abs/2007MNRAS.380.1573S} {380, 1573}

\bibitem[\protect\citeauthoryear{{Shafieloo}, {Alam}, {Sahni}  \&
  {Starobinsky}}{{Shafieloo} et~al.}{2006}]{2006MNRAS.366.1081S}
{Shafieloo} A.,  {Alam} U.,  {Sahni} V.,   {Starobinsky} A.~A.,  2006, \mn@doi
  [\mnras] {10.1111/j.1365-2966.2005.09911.x}, \href
  {http://adsabs.harvard.edu/abs/2006MNRAS.366.1081S} {366, 1081}

\bibitem[\protect\citeauthoryear{{Shafieloo}, {Kim}  \& {Linder}}{{Shafieloo}
  et~al.}{2013}]{2013PhRvD..87b3520S}
{Shafieloo} A.,  {Kim} A.~G.,   {Linder} E.~V.,  2013, \mn@doi [\prd]
  {10.1103/PhysRevD.87.023520}, \href
  {http://adsabs.harvard.edu/abs/2013PhRvD..87b3520S} {87, 023520}

\bibitem[\protect\citeauthoryear{{Sol{\`a}}, {G{\'o}mez-Valent}  \& {de Cruz
  P{\'e}rez}}{{Sol{\`a}} et~al.}{2017}]{2017MPLA...3250054S}
{Sol{\`a}} J.,  {G{\'o}mez-Valent} A.,   {de Cruz P{\'e}rez} J.,  2017, \mn@doi
  [Modern Physics Letters A] {10.1142/S0217732317500547}, \href
  {http://adsabs.harvard.edu/abs/2017MPLA...3250054S} {32, 1750054}

\bibitem[\protect\citeauthoryear{{Song} \& {Percival}}{{Song} \&
  {Percival}}{2009}]{2009JCAP...10..004S}
{Song} Y.-S.,  {Percival} W.~J.,  2009, \mn@doi [\jcap]
  {10.1088/1475-7516/2009/10/004}, \href
  {http://adsabs.harvard.edu/abs/2009JCAP...10..004S} {10, 004}

\bibitem[\protect\citeauthoryear{{Spergel} et~al.,}{{Spergel}
  et~al.}{2015}]{2015arXiv150303757S}
{Spergel} D.,  et~al., 2015, preprint, \href
  {http://adsabs.harvard.edu/abs/2015arXiv150303757S} {} (\mn@eprint {arXiv}
  {1503.03757})

\bibitem[\protect\citeauthoryear{{Weinberg}}{{Weinberg}}{1989}]{1989RvMP...61....1W}
{Weinberg} S.,  1989, \mn@doi [Reviews of Modern Physics]
  {10.1103/RevModPhys.61.1}, \href
  {http://adsabs.harvard.edu/abs/1989RvMP...61....1W} {61, 1}

\bibitem[\protect\citeauthoryear{{Zhao} et~al.,}{{Zhao}
  et~al.}{2017}]{2017NatAs...1..627Z}
{Zhao} G.-B.,  et~al., 2017, \mn@doi [Nature Astronomy]
  {10.1038/s41550-017-0216-z}, \href
  {http://adsabs.harvard.edu/abs/2017NatAs...1..627Z} {1, 627}

\bibitem[\protect\citeauthoryear{{de Felice} \& {Tsujikawa}}{{de Felice} \&
  {Tsujikawa}}{2010}]{2010LRR....13....3D}
{de Felice} A.,  {Tsujikawa} S.,  2010, \mn@doi [Living Reviews in Relativity]
  {10.12942/lrr-2010-3}, \href
  {http://adsabs.harvard.edu/abs/2010LRR....13....3D} {13}

\bibitem[\protect\citeauthoryear{{de la Torre} \& {Peacock}}{{de la Torre} \&
  {Peacock}}{2013}]{2013MNRAS.435..743D}
{de la Torre} S.,  {Peacock} J.~A.,  2013, \mn@doi [\mnras]
  {10.1093/mnras/stt1333}, \href
  {http://adsabs.harvard.edu/abs/2013MNRAS.435..743D} {435, 743}

\makeatother
\end{thebibliography}

\appendix

\section{Data visualization}

\begin{figure}
  \includegraphics[width=\columnwidth]{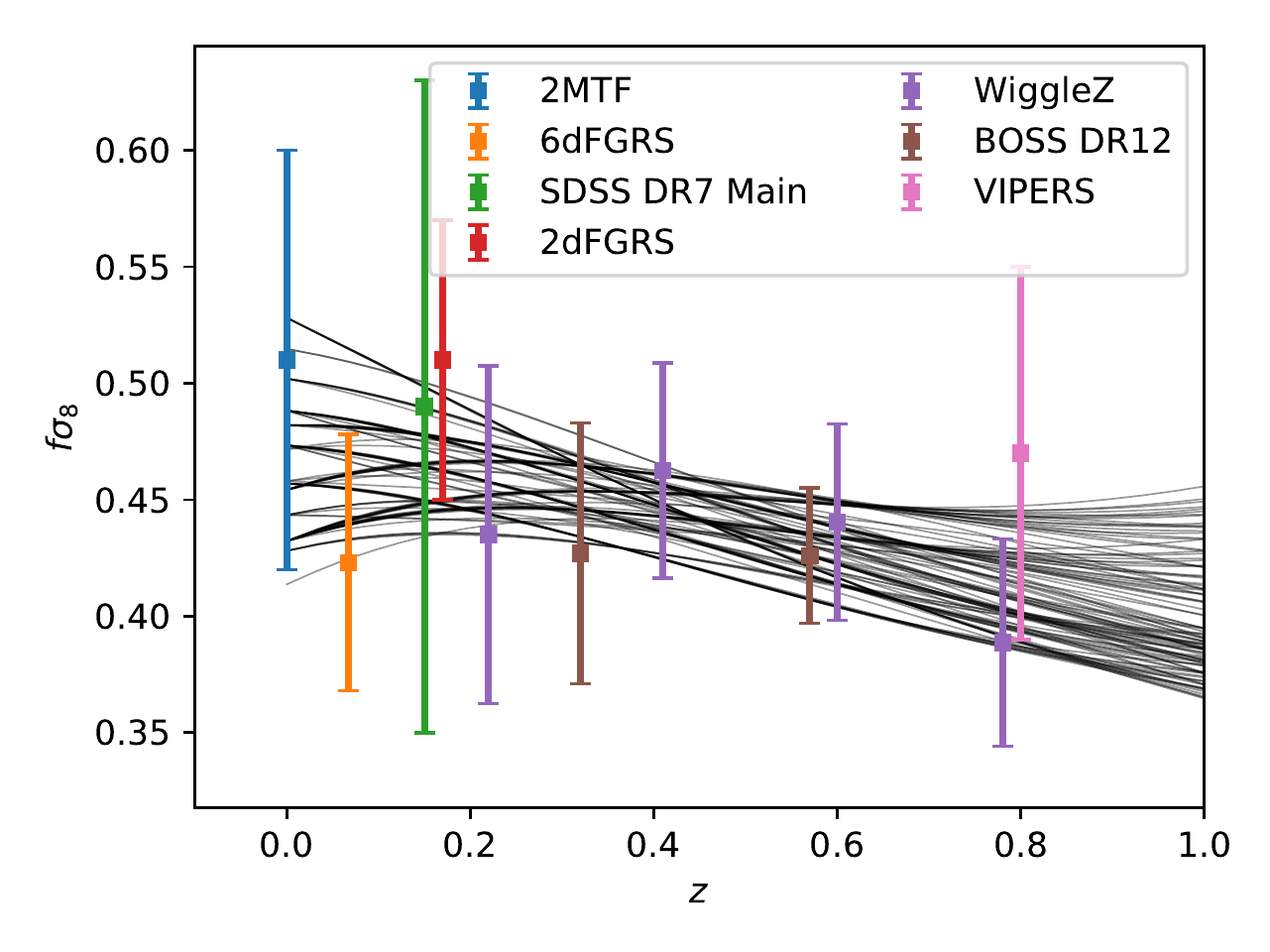}
  \caption{\label{fig:data_fs8}\fseight\   data    and   reconstructed
    \fseight\ with free $(\Omm,\gamma,\sigma_8)$.
    All lines shown here have
    $\chi^2<\chi^2_\text{\lcdm}$. }
\end{figure}

For   visualization   purpose,  Fig.~\ref{fig:data_fs8}   shows   some
reconstructed  $\fseight$  verifying    eq.~\eqref{eq:valid}  and  with
$\chi^2<\chi^2_\text{\lcdm}$. 

\bsp	
\label{lastpage}
\end{document}